\documentclass[11pt,fleqn,leqno]{article} 
\usepackage[final]{graphicx}
\usepackage[cmex10]{amsmath} 
\usepackage{amssymb}
\usepackage{amsfonts}
\usepackage{amsthm}
\usepackage{setspace}
\usepackage{epstopdf}
\usepackage{natbib}
\usepackage[affil-it]{authblk}
\usepackage[final,colorlinks]{hyperref}
\usepackage[margin=1in]{geometry}
\usepackage{longtable}

\newcommand{\abs}[1]{\left\lvert#1\right\rvert}

\begin{document}
\title{Optimal timing of cross-sectional network samples in longitudinal network studies}
\author[$\dag$]{Ekkehard Beck\footnote{Email addresses: ebeck@u.northwestern.edu (Ekkehard Beck), 
armbrusterb@gmail.com (Benjamin Armbruster)}}
\author[ ]{Benjamin Armbruster}
\affil[$\dag$]{Department of Industrial Engineering and Management Sciences,
Northwestern University, Evanston, IL, 60208, USA}
\date{\today}

\maketitle

\begin{abstract}
When choosing the timing of cross-sectional network snapshots in longitudinal social network studies, the effect on the precision of parameter estimates generally plays a minor role.  Often the timing is opportunistic or determined by a variety of considerations such as organizational constraints, funding, and availability of study participants.  Theory to guide the timing of network snapshots is also missing.  We use a statistical framework to relate the timing to the precision of the parameter estimates, specifically, the sum of the relative widths of their confidence intervals. We illustrate this computationally using the STERGM suite of the \emph{statnet} package to estimate the parameters of the network dynamics.  Analytically, we derive simple approximations for the optimal timing when the parameters correspond to the rates for different network events.  We find that the optimal time depends the most on the network processes with short expected durations and few expected events such as the dissolution of ties. We apply our approximations to a simple example of a dynamic network with formation and dissolution of iid ties. 
\end{abstract}


\section{Introduction}

Beginning in the 1960's, longitudinal social network studies such as Sampson's monastery study \citep{Sampson1969} or Newcomb's fraternity friendship network study \citep{Newcomb1961} and their analyses have led to an increased understanding of complex interactions among individuals across a wide range of disciplines such as sociology \citep{Newman2001,Steglich2010}, communication and marketing science \citep{Onnela2007}, computer science \citep{Kuhn2010}, neuroscience \citep{Park2013}, cell biology \citep{Han2004}, epidemics \citep{Morris1995}, and ecology \citep{Pascual2006}. Most of the theoretical and methodological work in these disciplines focuses on the visualization and statistical analysis of already sampled dynamic network data, specifically the inference of network dynamics, network topology, and the mechanisms driving these dynamics \citep{Holme2012,Holme2015}.  Few works consider the sourcing of dynamic network data and the actual sampling of social network data over time \citep{Caceres2013}.

Often sampling in longitudinal social network and clinical cohort studies is done opportunistically \citep{Caceres2013} and the decision when to sample is often dependent on a wide range of factors \citep{Timmons2015} such as the type of social network (e.g., email correspondence vs. sexual partnership network); competing objectives, processes, and measurements (e.g., a social network study nested inside a clinical cohort study \citep{Kuhns2015,Mustanski2014,Mustanski2015}); the available time or planned time horizon of the study; the availability of funds, logistical or organizational constraints; the availability of study participants; the expected behavior of the study participants (i.e., attrition, task adoption, and task adaptation of study participants); as well as the desired accuracy and precision of the expected results \citep{Timmons2015}. Often factors which only indirectly influence the measurement of effects are given more consideration than those which directly relate to the measurement of effects such as the reliability of the expected study results (i.e., accuracy and precision). Focusing on accuracy and precision would help prevent the under- or overestimation of effects as well as the inability to detect effects resulting from time-dependent correlations and time-dependent regression coefficients, and ultimately help prevent inaccurate and potentially misleading conclusions \citep{Timmons2015,Butts2009}. 

Collecting cross-sectional networks or network snapshots at discrete points in time is one of the most important sampling designs and representations for dynamic network data in longitudinal social network and clinical cohort studies with social network data \cite{Robins2015}.  However, most such  studies do not discuss how they decided various aspects of the temporal sampling design such as their choice of the timing between network samples and the number of waves sampled. Rather, the majority of such studies only state the outcome of their decision making such as \citet{Knecht2006} in her study of friendship selection and friends' influence in Dutch schools where the timing between waves is discussed by ``in each school year four waves were scheduled with three months in between consecutive waves.'' When studies do provide some rationales for their choices, these are most often not motivated by accuracy and precision arguments, such as in the case of a study by \citet{Lubbers2010} who study the acculturation of immigrants in south Florida and northeastern Spain. There, the authors state that the time duration of two years between two samples resulted from two consecutive grant fundings  \citep{Lubbers2010,Mccarty2004}. In another example, \citet{Mercken2012} state in their nested study about peer influence, peer selection, and smoking behavior among British adolescents that the timing between cross-sectional network samples was determined by the design of the underlying randomized clinical trial studying the effects of a peer-based smoking intervention over time.

Among the theoretical and methodological approaches focusing on sampling of dynamic network data and the representation of dynamic networks, most focus on the aggregation of links in time-window graphs \citep{Holme2012,Holme2015}. Here, all links which are present during a specified observation period are aggregated to a static or cross-sectional network graph. Because time-windows are finite, they are likely to begin or end during inter-event times of dynamic network processes such as the formation and dissolution of ties. Thus, the choice of the finite observation period or time-window impacts the observation bias and therefore influences the accuracy and significance of the analysis and inference of the observed dynamic network \citep{Holme2012,Holme2015,Kivelae2015}. \Citet{Fish2015} address this question when studying the relationship between the length of time-windows used to aggregate observed network ties over time and the temporal scale of network processes. Using link prediction (i.e., to predict the occurrence of ties in the future based on past network snapshots) as an outcome variable and considering the inherent noise in an aggregated network snapshot due to oversampling (i.e., the noise is dependent on the length of the time-window), Fish and Caceres find that link prediction performs better on sequences aggregated at window sizes close to the ground-truth than on other window sizes. Further, the findings of \citet{Psorakis2012} who study spatio-temporal data sets of wild birds using a Gaussian mixture model for event-streams, suggest that for short window-sizes the actual starting time in case of periodic network dynamics has a significant impact on the observation bias besides the length of the chosen window-time. 

Despite these methodological approaches to determine the optimal choice and timing of time-windows to aggregate already sampled data, we are not aware of any work in the dynamic social network analysis literature which considers measurement precision when determining the time between consecutive cross-sectional network samples. Thus, we extend our review to literature about the optimal timing of consecutive samples in longitudinal studies in psychology, medicine, and digital signal processing. In the field of digital signal processing, Shannon's theorem \citep{Shannon1949} states that the sampling frequency should be at least twice the highest frequency in the signal. However, the applicability of this theorem is unclear since social network studies are rarely interested in periodic behavior. For randomized clinical trials,  \citet{Wilkens2005} derives the optimal timing between two repeated measurements for linear mixed effects model with two treatment effect patterns in order to maximize the efficiency in estimating the treatment effects.
Despite these and related theoretical approaches, most discussion in psychology and the social sciences of sampling design refers to the rule of thumb that the more often one samples the better \citep{Timmons2015,Collins2002,Siegler2006}. Using computational experiments, \citet{Timmons2015} study the impact of the time between two measurements and the unequal spacing 
across consecutive samples on the precision of estimates. While sampling more often helps, Timmons and Preacher observe diminishing returns and thus suggest that researchers consider the tradeoff of precision versus sampling time.

In this study, we place the problem of when to collect cross-sectional network snapshots in the framework of statistical study design.  We use this framework to relate the time between consecutively sampled cross-sectional network snapshots and the precision of parameter estimates derived from these snapshots. We then derive easy-to-use approximations for the optimal timing between consecutive samples. We hope that this first step provides researchers designing longitudinal social network or clinical cohort studies useful insight to increase the precision of their findings.

Section~\ref{sec2} presents the statistical framework, derives the analytic approximations, and then applies them to a continuous time Markov Chain (CTMC) based dynamic network model with formation and dissolution of independent and identically distributed (iid) ties. In section~\ref{sec3}, we simulate a dynamic network over time.  We use this to compare our analytic approximations for the optimal spacing with computational experiments where we use the STERGM \citep{Krivitsky2014STERGM} package of the \emph{statnet} suite \citep{statnet} to estimate the parameters of the network dynamics for various timings of two network observations. We conclude in section~\ref{sec4}.

\section{Statistical framework and analytics}\label{sec2}

\subsection{Statistical framework}\label{sec2.1}
Our first contribution is to frame the problem in terms of statistical study design.  Suppose we have parameters $\vec{\alpha}:=(\alpha_1,\dotsc,\alpha_m)$ which we hope to estimate.  The standard approach of decision theory is to find an action $a$ that maximizes the expectation of some utility $E_{\alpha}[u(a)]$, where the distribution depends on the parameters.  Similarly, the standard statistical approach is to find estimates $\hat{\alpha}$ where the expectation of the loss function,  $E_{\alpha}[L(\vec{\alpha},\hat{\alpha})]$, is small (a common loss function is the mean-squared error).  The goal of statistical study design is to choose the design parameters $h$ (in our case when to collect the data) which improves the expected utility or loss the most.  In a Bayesian decision theory, we would choose $h$ to maximize the expected ``value of information''.

We will not use a Bayesian or decision theory approach since that requires choosing a prior distribution and a utility function.  Instead for simplicity, the value we seek to minimize will be the expected sum of the relative widths of the confidence intervals (CI) of the parameters.  Let $\hat{\alpha}_i$ be our point estimate for $\alpha_i$ and $[\alpha_{i,-},\alpha_{i,+}]$ the CI.  We then define 
\begin{equation}
	w_i(h):= E[(\alpha_{i,+} - \alpha_{i,-})/\abs{\hat{\alpha}_i}]
\end{equation}
to be the expectation of the relative width of the CI (i.e., the ratio of the width of the CI to the absolute value of the point estimate).  Then our goal is to minimize
\begin{equation}
	\min_h w_1(h)+\cdots+w_m(h).
\end{equation}
If some of parameters are more important than others, then we can take a multi-objective optimization approach, for example by weighting the $w_i(h)$ terms.

\subsection{Analytics}\label{sec2.2}
Our second contribution is to develop useful approximations of the optimal time between snapshots.  In this section we develop the analysis.  First we consider a single timer type, with an exponential distribution, (for say the duration of a relationship) and relate our censored data (whether or not the relationship still exists at the second snapshot) to the precision of our estimate for the expected duration.  Then we consider multiple timer types with exponential distributions (e.g., two types for edge formation and edge dissolution) and determine the measurement interval between two cross-sectional network samples which minimize the combined measurement uncertainty among the parameters to be estimated. Finally, we apply this framework to a simple example of a CTMC-based dynamic network with iid formation and dissolution of ties.

\subsubsection{A single timer type}
Suppose we have a nonnegative (timer) random variable, $X(\alpha)$, parameterized by $\alpha$ that represents the duration of some condition (e.g., how long a certain edge exists). After time $h$, we check if the condition still holds, that is we observe $Y(\alpha)=I[X(\alpha)\le h]$, where $I[\cdot]$ is an indicator random variable.  Here $Y$ is essentially a censored observation of $X$.  Suppose we have $n$ iid timers $X_1,\dotsc,X_n$ giving observations $Y_1,\dotsc,Y_n$.
Our goal is to estimate $\alpha$ from these observations. We let $w(h;n,\alpha)$ denote the precision of our estimate, specifically the expected relative width of the CI of our estimate of $\alpha$. Thus, our goal is to find the $h$ minimizing $w(h;n,\alpha)$:
\begin{equation}
	\min_h w(h;n,\alpha).
\end{equation}

\subsubsection{Multiple timer types}
We now consider multiple timer types (e.g., the duration of an edge and the lifespan of a node). Suppose there are $m$ types of timers with parameters $\alpha_i$ for $i=1,\dotsc,m$, respectively.  For type $i$, we have $n(i)$ iid timers, $X_{i,1},\dotsc,X_{i,n(i)}$ giving observations $Y_{i,1},\dotsc, Y_{n(i),1}$. Generally, there will not be a single value of $h$ that estimates all the parameters with optimal precision.  Our goal is to choose an $h$, which gives a good tradeoff among the precision of the estimates of the various parameters.  We use the following objective function,
\begin{equation}
\min_h \sum_i^m w(h;n(i),\alpha_i).
\end{equation}

\subsubsection{Deriving the objective function}
Let $p=P[Y=1]$ be the true success probability and $\hat{p}=(\sum_i^m Y_i)/n$, the empirical point estimate. Let $[p_{-}, p_{+}]$ be the CI with confidence level $1-\kappa$. Since $p(\alpha;h)=P[X(\alpha)\le h]$ we are able to transform the point estimate and CI for $p$ into a point estimate and CI for $\alpha$.

We assume $X(\alpha)$ to be an exponentially distributed random variable.  Hence, $p=1-\exp(-h\alpha)$ and thus, $\alpha=(-1/h)\log(1-p)$. We then have the following point estimate and CI for $\alpha$,
\begin{align}
	\hat{\alpha} &= (-1/h)\log(1-\hat{p}),\\
	[\alpha_-, \alpha_+] &= (-1/h)[\log(1-p_{-}),\log(1-p_{+})].
\end{align}
Therefore, the relative width of the expected confidence interval is 
\begin{equation}\label{eq:wObj}
w(h;n,\alpha)=(1/\alpha)(-1/h)E[\log(1-p_{+})-\log(1-p_{-})].
\end{equation}

\subsubsection{Approximating the objective function}
We can now numerically evaluate \eqref{eq:wObj} using any of the common Binomial confidence intervals \citep{Brown2002}.
We will use the normal approximation interval, where $p_{\pm}$ is $\hat{p}\pm (z/ \sqrt{n})\sqrt{\hat{p}(1-\hat{p})}$, where $z$ is the $z$-score factor (e.g., 1.96 for a $95\%$ confidence interval), and thus obtain
\begin{multline}\label{eq:wObj2}
w(h;n,\alpha)=(1/\alpha)(-1/h)E[\log(1-\hat{p}-(z/ \sqrt{n})\sqrt{\hat{p}(1-\hat{p})})\\
- \log(1-\hat{p}+(z/ \sqrt{n})\sqrt{\hat{p}(1-\hat{p})})].
\end{multline}

We simplify \eqref{eq:wObj2}  by replacing the stochastic $\hat{p}$, by the deterministic $p$ (i.e., replacing the sample  estimate with the true population mean, which again assumes that we have high accuracy in our measurements),    
\begin{multline}\label{eq:wApprox}
w(h;n,\alpha)\approx -(\alpha h)^{-1}[\log(1-p-(z/ \sqrt{n})\sqrt{p(1-p)})\\
- \log(1-p+(z/ \sqrt{n})\sqrt{p(1-p)})],
\end{multline}
where $p$ is a function of $\alpha$ and $h$ as defined above.

However, to obtain an analytically tractable solution we have to further approximate. For large $n$, we can linearize around $\log(1-p)$, i.e.,
\begin{equation}
\log(1-p+\epsilon)\approx \log (1-p)+\epsilon/(1-p),
\end{equation}
and thus obtain,
\begin{equation}
w(h;n,\alpha)\approx (\alpha h)^{-1}2(z/ \sqrt{n})\sqrt{p(1-p)}/(1-p)
=2z/(h \alpha \sqrt{n})\sqrt{p/(1-p)}.
\end{equation}
Substituting back $p=1-\exp(-h\alpha)$, 
\begin{equation}\label{eq:wSingleTimer}
w(h;n,\alpha) \approx \frac{2z}{\sqrt{n}} \frac{\sqrt{\exp(h\alpha)-1}}{h\alpha}.
\end{equation}
As expected from the central limit theorem, the relative width of the confidence interval scales with $n^{-1/2}$.

\subsubsection{Optimal spacing}
We determine numerically that $\frac{\sqrt{\exp(x) -1}}{x}$ has a minimum at $\gamma_0\approx 1.59$. Thus, for a single timer type, the optimum observation time is $h^*\approx \gamma_0/ \alpha$, which supports our intuition that in this case the optimal measurement interval is a constant multiple of the expected time duration.

For multiple timers, we apply the Taylor expansion around $x=\gamma_0$,
\[ \frac{\sqrt{\exp(x) -1}}{x} \approx \gamma_1+\gamma_2(x-\gamma_0)^2, \]
where $\gamma_1$ and $\gamma_2$ are constants, to further simplify \eqref{eq:wSingleTimer}:
\begin{equation}
w(h;n,\alpha) \approx 2zn^{-1/2}(\gamma_1+\gamma_2(h\alpha-\gamma_0)^2).
\end{equation}
Then our objective function is approximated as,
\begin{equation}
\sum_i w(h;n(i),\alpha_i) \approx 2z\sum_i^m n(i)^{-1/2}(\gamma_1+\gamma_2(h\alpha_i-\gamma_0)^2),
\end{equation}
which has a first order condition,
\begin{equation}
0=\sum_i^m n(i)^{-1/2}\alpha_i(\alpha_i h-\gamma_0)=h\sum_i^m \alpha_i^2 n(i)^{-1/2}-\gamma_0 \sum_i^m \alpha_i n(i) ^{-1/2},
\end{equation}
and thus the optimum is
\begin{equation}\label{eq:hstar1}
h^*\approx \gamma_0 \Bigl(\sum_i^m \alpha_i n(i)^{-1/2}\Bigr)/ \Bigl(\sum_i^m \alpha_i^2 n(i)^{-1/2}\Bigr).
\end{equation}

Thus, the reciprocal of the optimum observation time, $1/h^*$, is an average of the optimal values for each timer type weighted by $\alpha_i n(i)^ {-1/2}$, giving priority to timers with small expected durations, $1/\alpha_i$, and small number of samples, $n(i)$:
\begin{equation}\label{eq:hstar2}
1/h^*\approx \sum_i^m \frac{\alpha_i}{\gamma_0}\frac{\alpha_i n(i)^{-1/2}}{\sum_i^m \alpha_i n(i)^{-1/2}}.
\end{equation}

\subsection{Example: dynamic network with formation and dissolution of iid ties}\label{sec2.3}

To illustrate our framework we apply it to a stylized  longitudinal network study with two cross-sectional network samples. Suppose that the first wave of the study is at time 0 and the second at $h$; that the study has $n$ nodes; and that the average degree at time 0 is $k$.  We assume that the network dynamics are very simple, that the dissolution and formation of undirected ties between any pair of nodes is independent and occurs at rates $d$ and $f$, respectively.  Thus, we have two timer-types, $m=2$: $n_d=nk/2$ dissolution timers with $\alpha_d=d$ and $n_f=n(n-1)/2-n_d$ formation timers with $\alpha_f=f$.

We seek to determine the optimal time $h^*$, which gives us the most precise estimates for the rates of the network process. Specifically, we seek to optimize $w_k+w_d$, the sum of the relative widths of the $95\%$ confidence intervals for $f$ and $d$. We also make the assumption that no node pair changes state more than once in that time period, that is we do not have an edge appearing between a pair of nodes and then subsequently disappearing between the two waves of the study.

\begin{figure}[h]
\centering
\includegraphics[scale=0.5]{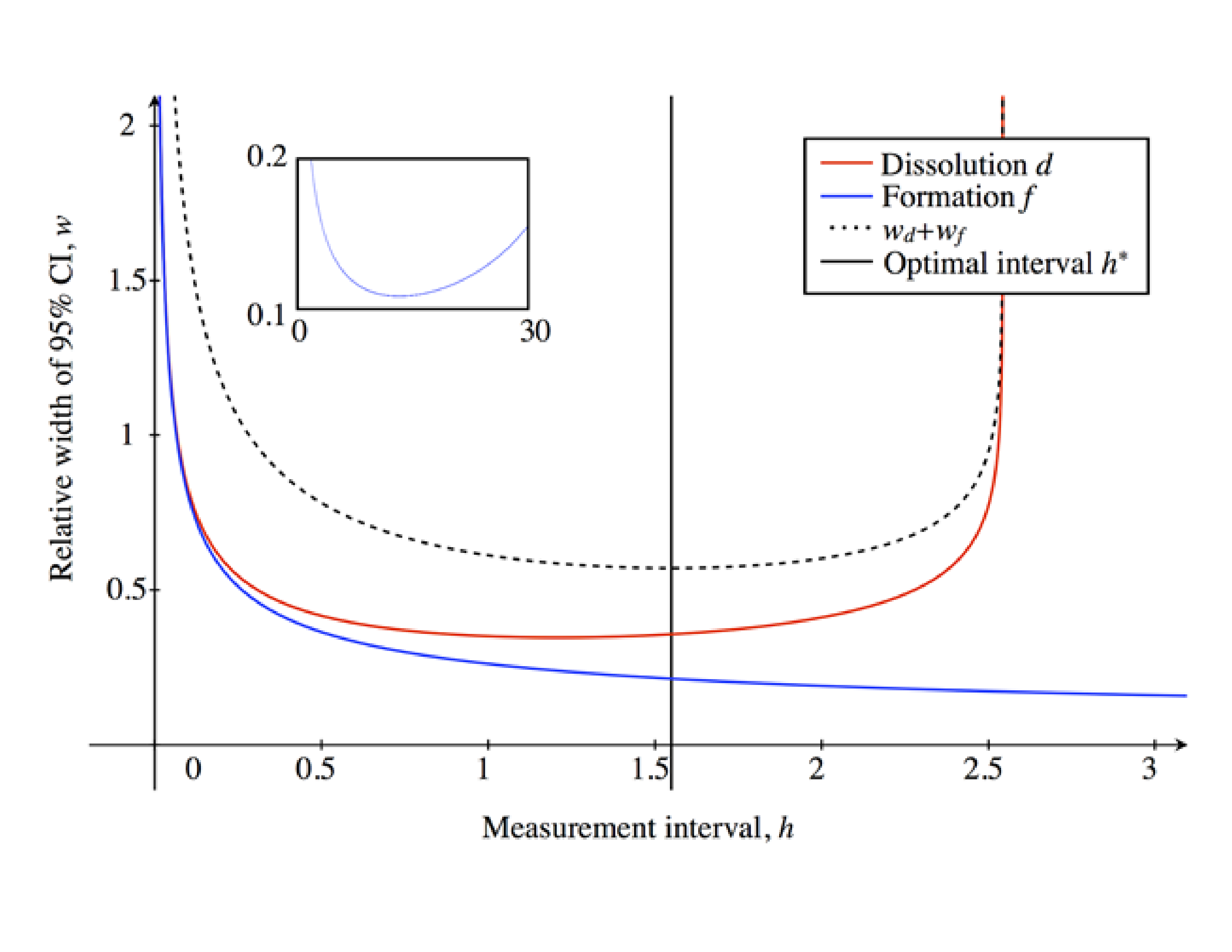}
\caption[Relative width of $95\%$ confidence intervals (CI) in a dynamic network with $n=30$ nodes]{The relative width of the $95\%$ confidence interval (CI) for the dissolution and formation rates, $d$ and $f$ in a dynamic network with iid ties.  The network has $n=30$ nodes, average degree $k=3$, $d=1$, and since we are assuming the network dynamics are in  steady-state, $f=dk/(n-1-k)\approx 0.12$.  The red and blue curves are calculated using the approximation \eqref{eq:wApprox}.  The dashed black line is the sum of the relative widths of the CI for the formation and dissolution rates.  The solid black line is the optimal measurement interval.  The inset shows how the blue curve eventually goes back up.}
\label{fig2.1}
\end{figure}

\begin{figure}[h]
\centering
\includegraphics[scale=0.5]{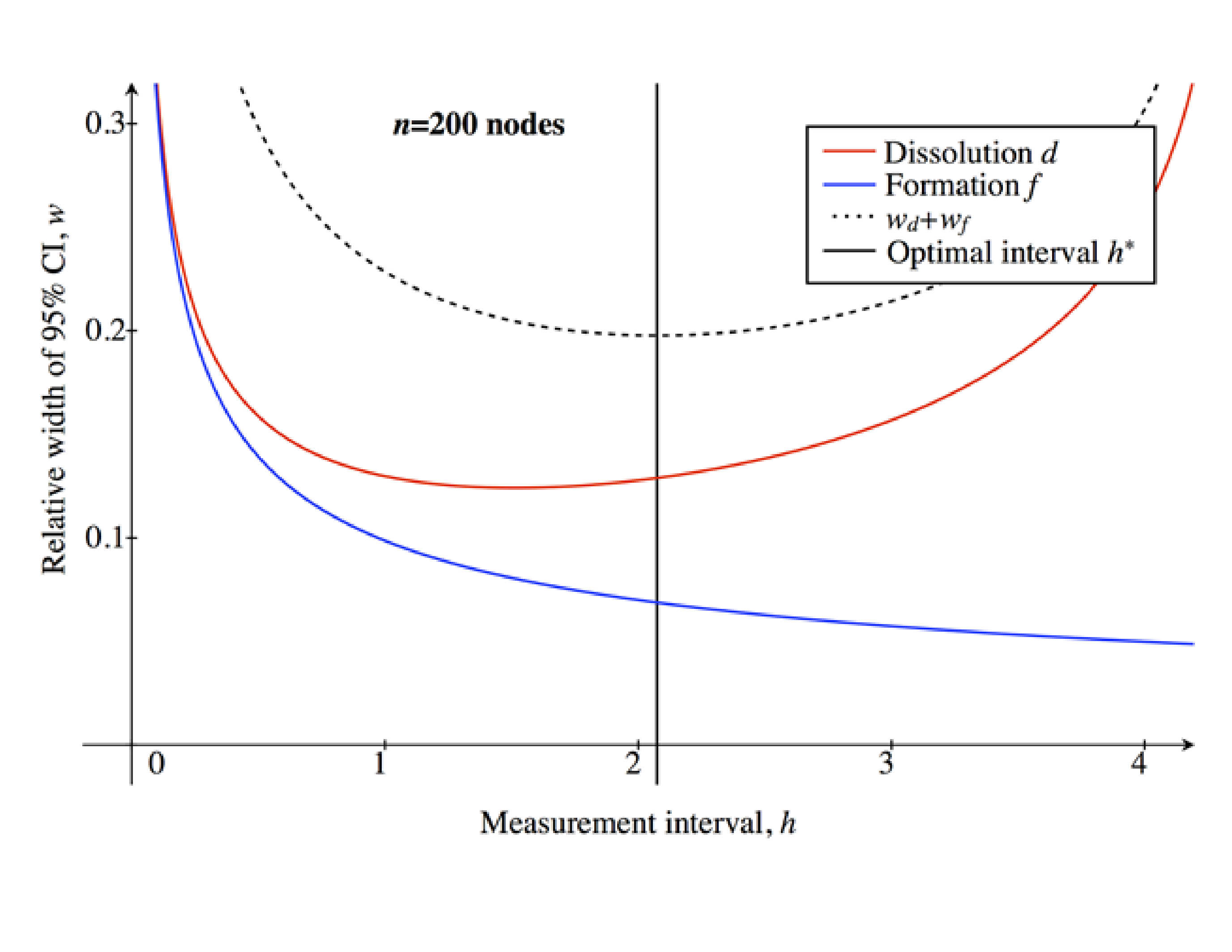}
\caption[Relative width of $95\%$ confidence intervals (CI) in a dynamic network with $n=200$ nodes]{The relative width of $95\%$ confidence intervals (CI) in a dynamic network with $n=200$ nodes.  The parameters are the same as in Figure \ref{fig2.1} except with $n=200$ nodes and thus due to the steady-state assumption, $f\approx 0.015$.}
\label{fig2.2}
\end{figure}

\begin{figure}[h]
\centering
\includegraphics[scale=0.5]{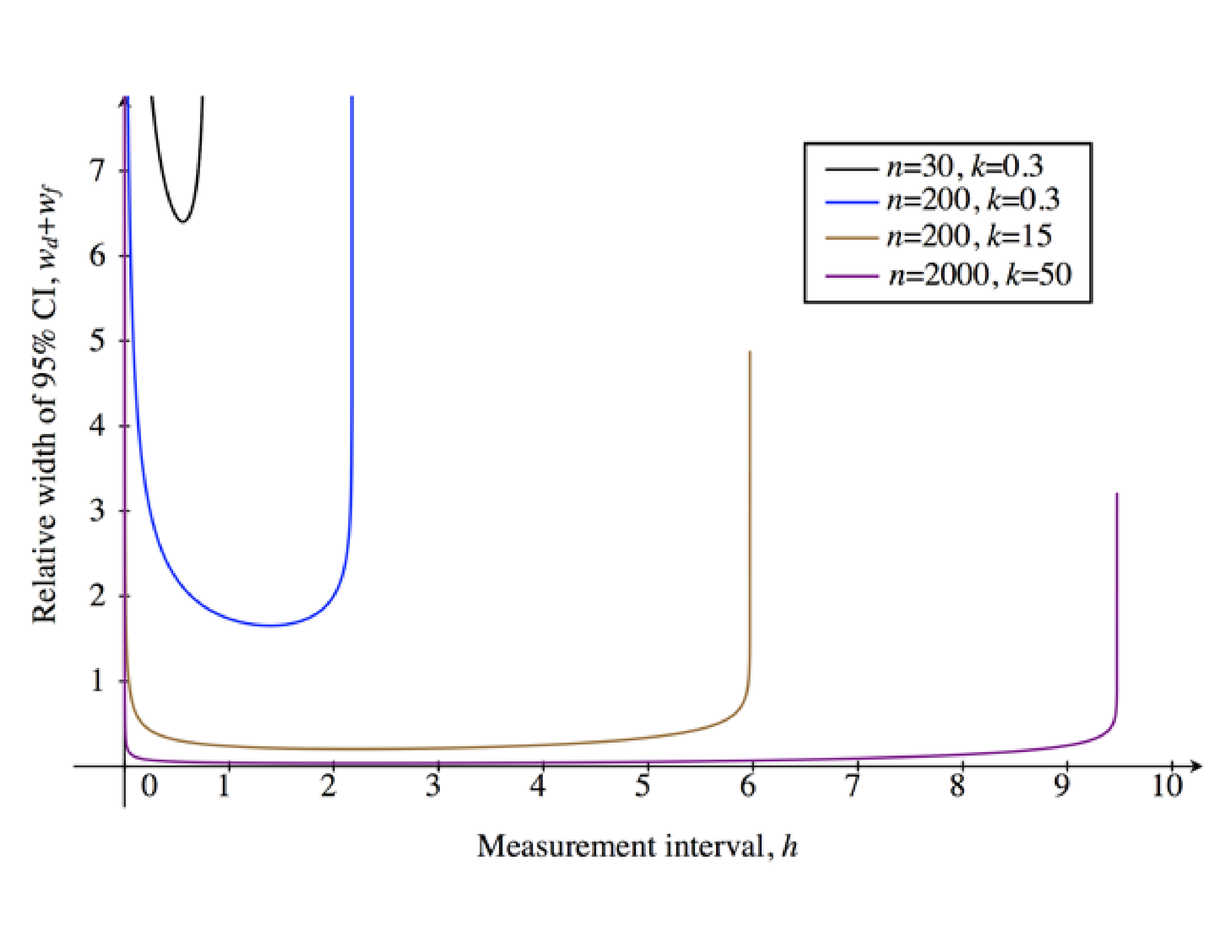}
\caption[Estimate precision for various values of $n$ and $k]{Precision of the formation and dissolution rates for various values of $n$ and $k$.  The precision is measured as the sum of the relative widths of the 95\% confidence intervals, calculated using approximation \eqref{eq:wApprox}.}
\label{fig2.3}
\end{figure}

Figure \ref{fig2.1} shows the precision of the estimates for $d$ and $f$ as a function of $h$ for a case with $n=30$ nodes.  The vertical asymptote of $w_d$, the red line, on the right is due to the normal approximation of the confidence interval, which breaks down if $p$ is close to 0. This is not a problem since we care about values close to the minimum, which are away from the asymptote.  Figure \ref{fig2.2} uses the same parameter values as Figure \ref{fig2.1}, but with $n=200$ nodes. Figure \ref{fig2.3} considers further parameter values. 

Note that the optimal measurement interval, $h^*$, in these figures (calculated using approximation \eqref{eq:wApprox}), is of the same order of magnitude as $\gamma_0/d$, the approximate measurement time if we were only interested in the dissolution rate.  (In contrast, the expected time until an edge forms between two nodes, $1/f\approx n/k$, is much larger.)
In fact, for all but the case of $n=30$ and $k=0.3$, $\gamma_0/d$ is close to optimal and part of the relatively flat portion of the objective.  This suggests that we should only focus on the dissolution rate when choosing our measurement time.
Analytically, we can justify that $h^*$ is close to $\gamma_0/d$ using 
\eqref{eq:hstar1},
\begin{equation}
h^* \approx \gamma_0 \frac{\frac{f}{\sqrt{n-1-k}}+\frac{d}{\sqrt{k}}}{\frac{f^2}{\sqrt{n-1-k}}+\frac{d^2}{\sqrt{k}}}, 
\end{equation}
and the fact that in many contexts, $n-1-k\gg k$ and $d\gg f$, since there are generally many fewer edges in a network than pairs of nodes without edges between them.

\section{Computational experiments}\label{sec3}

We now follow up our previous section on theory with a section on computational experiments.  
We simulate the evolution of two undirected dynamic networks in continuous time using the algorithm of \citet{Snijders2013} where the time between subsequent events is exponentially distributed with rate $\rho$. At each event, a pair of nodes is chosen at random and an edge between them is set (irrespective of whether an edge currently exists) based on the probability that this edge exist in an exponential random graph model (ERGM) conditional on all the other edges in the network (see equation 11.2 in \citet{Snijders2013}).  
Over time the network will reach a steady-state where it will be distributed according to an ERGM distribution with those same parameters. Table~\ref{table1} shows the parameters for our two dynamic networks: a simple and a more complex simulation.  The simple dynamic network has only a single ERGM parameter and is equivalent to the example with iid edges we had in section~\ref{sec2.3}.  In the more complex simulation, the probability that an edge forms or dissolves also depends on the neighbors of the edge's endpoints via the two-path, triangle, and three-path coefficients.

\begin{longtable}{|p{5cm}p{4.5cm}p{4.5cm}|}
\caption[Parameters of the dynamic network simulations]{Parameters of the dynamic network simulations using the algorithm of \citet{Snijders2013}.}
\label{table1}\\
\hline {\textbf{Parameters}} & {\textbf{Simple dynamic}} & \textbf{More complex}\\  
  & {\textbf{(network)}} & \textbf{(dynamic network)}  \\  
\hline
\endfirsthead
\hline 
\endlastfoot
Simulation time, $T$ & 10,000 & 10,000\\
Burn-in time, $t_{\mathrm{burn-in}}$ & 5000 & 5000\\
Replications & $n_{\mathrm{E,simple}}=250$ & $n_{\mathrm{E,complex}}=100$\\
\hline
 Number of nodes, $n$& 100 & 100 \\ 
 Intensity parameter, $\rho$ & 15 & 30  \\ \hline
 ERGM parameters& edge -3.89 & edge -2.5\\ 
 & & two-path -0.5 \\
 & &  triangle 2.0\\
 & & three-path 2.0
\end{longtable}

In this approach using ERGM probabilities, the formation and dissolution rates are not independent (but depend on shared network statistics).  Thus the analysis based on independent timers of multiple types in sections~\ref{sec2.2} and \ref{sec2.3} is not really justified here and we base our approach on the more general framework in section \ref{sec2.1}, specifically the goal of minimizing the sum of the CI of the parameters being estimated.  Never the less, we will use the simple approximation for $h^*$ from section~\ref{sec2.2} as a reference.

Starting with an empty network, we simulate the network until time $T=10,000$.  We assume that after time $t_{\mathrm{burn-in}}=5000$ the network is in steady-state.  Then for various values of $h$, we take two network snapshots after $t_{\mathrm{burn-in}}$ that are spaced $h$ apart and estimate the ERGM parameters.  
The estimation is done using the STERGM package \citep{Krivitsky2014STERGM} of the \emph{statnet} suite \citep{statnet} implemented in \emph{R} \citep{R}.  
Each simulation and estimation is replicated $n_{\mathrm{E,simple}}=250$ and $n_{\mathrm{E,complex}}=100$ times, respectively, and we only count accurately fitted models (i.e., with non-degenerate fits).
These replications allow us to calculate 95\% CIs empirically using the normal approximation.  The spikes in the relative CI size are due either to the software having difficulty finding a stable estimate or to the sample mean of the estimates being close to 0, leading to the \emph{relative} widths of the CI being large.  We did not explore these artifacts further since our computations are mainly for illustrative purposes.

\begin{figure}[h]
\centering
\includegraphics[width=\textwidth]{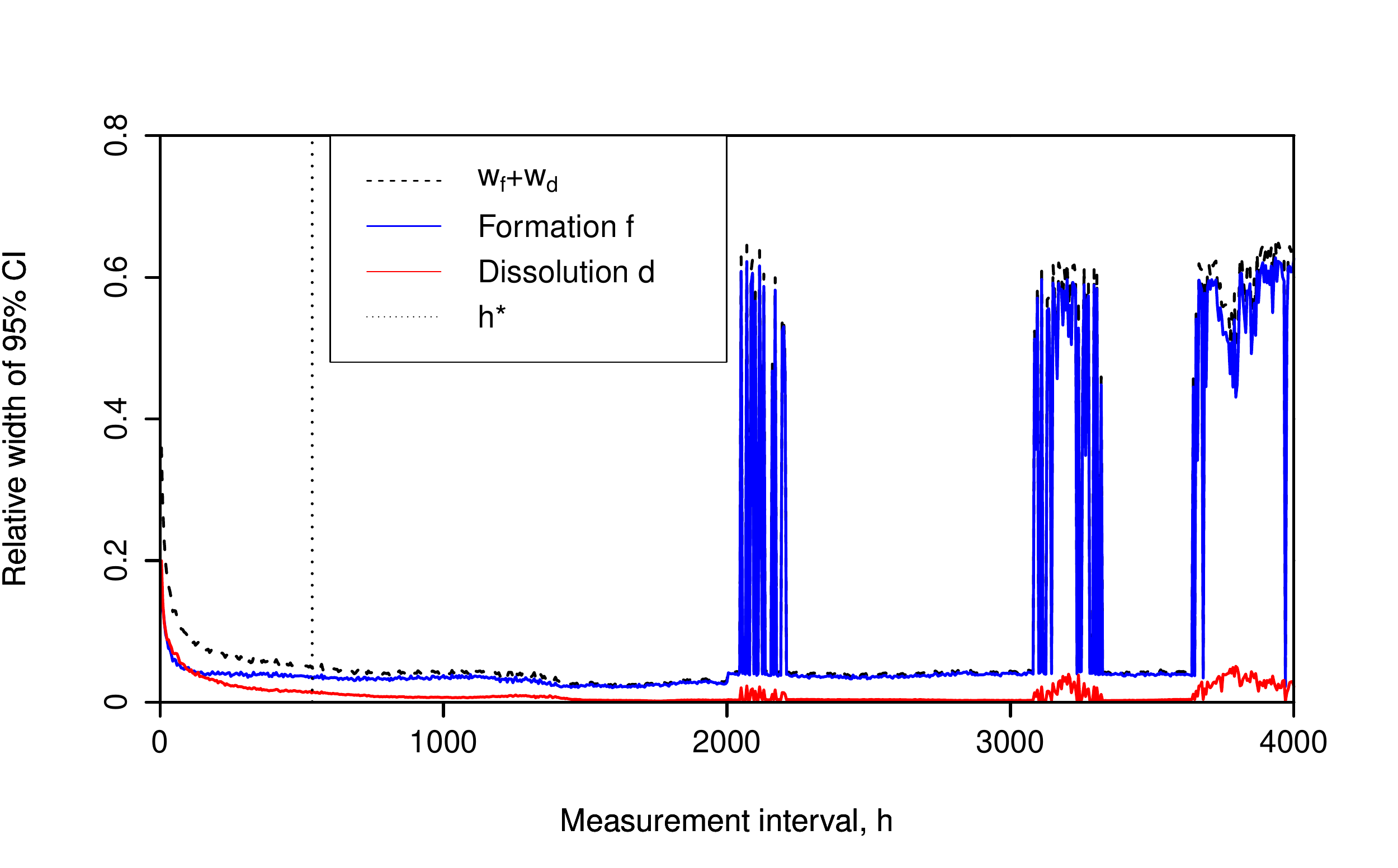}
\caption[Precision of the formulation and dissolution rates for the independent edges model.]{The relative width of the $95\%$ confidence interval (CI) for the dissolution and formation rates in the simple dynamic network.  We also show the sum of the relative widths, and the dashed vertical line at $h^*$ shows the analytically derived optimum measurement interval (see section \ref{sec2.2}).  The coefficients are estimated using the STERGM package \citep{Krivitsky2014STERGM} of the \emph{statnet} suite \citep{statnet} from two network snapshots spaced $h$ time units apart.  The formation and dissolution rates were derived from the coefficients according to \citet{Snijders2013}.  The CI are calculated empirically using the normal approximation from the 250 replications.}
\label{fig4}
\end{figure}

For the simple dynamic network, we can calculate formation and dissolution rates from the estimated coefficients using the Markov generator of the simulation process \citep{Snijders2013}.  This allows us to create Figure~\ref{fig4} analogous to Figure~\ref{fig2.1} showing the relative CI of both the formation and the dissolution rates.
We calculate that the optimum measurement interval $h^*\approx 537$ using the approximation from section~\ref{sec2.2}.

Figure~\ref{fig:appendix} in the appendix shows the actual average formation and dissolution rates and coefficients for the simple dynamic network.  From it we see that too short a measurement interval will bias the derived formation and dissolution rates towards zero, likely because few formation and dissolution events are observed in the measurement interval.  It is reassuring that the theoretical approximation for $h^*$ is close to when the estimates stabilize.

\begin{figure}[h]
\centering
\includegraphics[width=\textwidth]{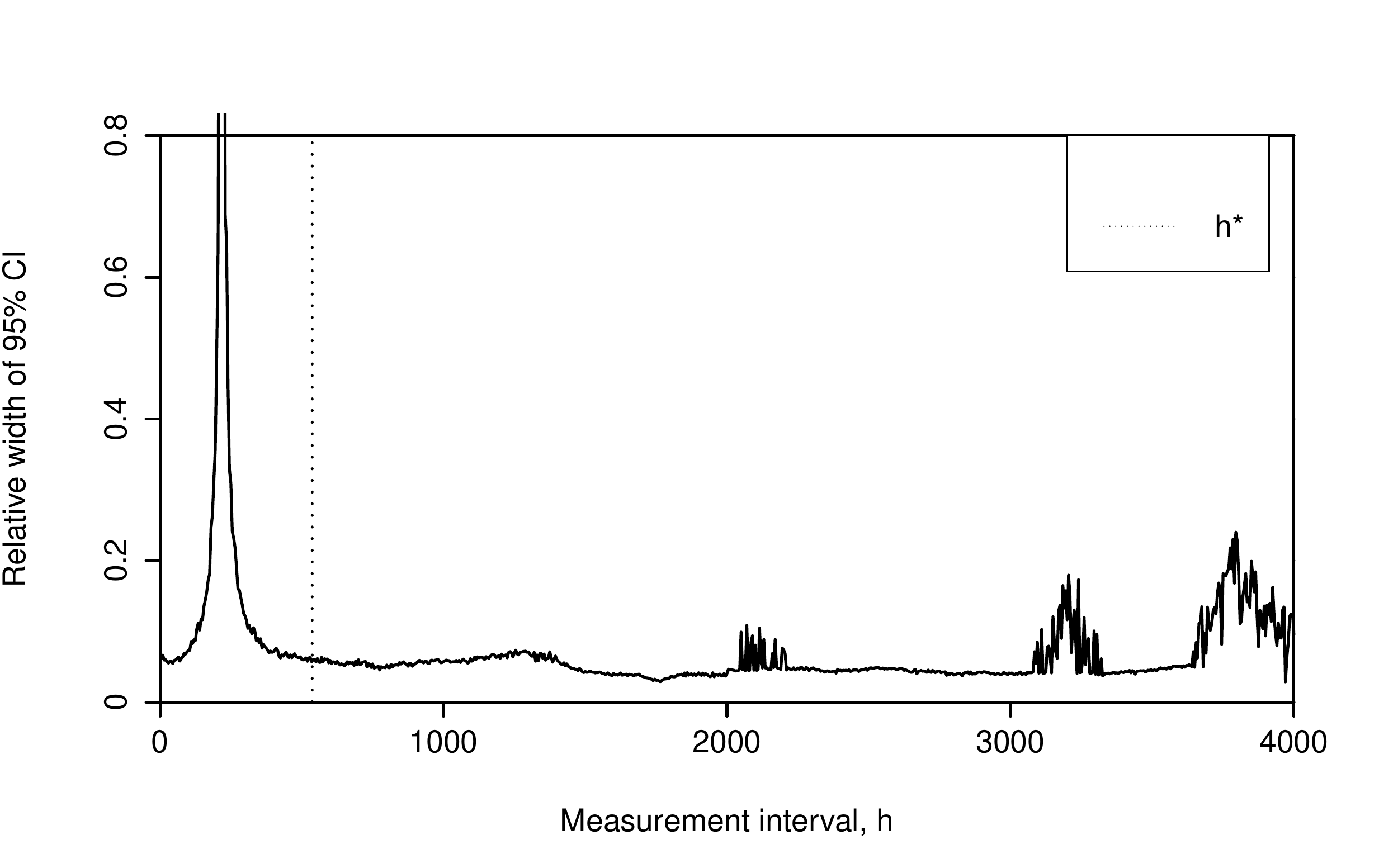}
\includegraphics[width=\textwidth]{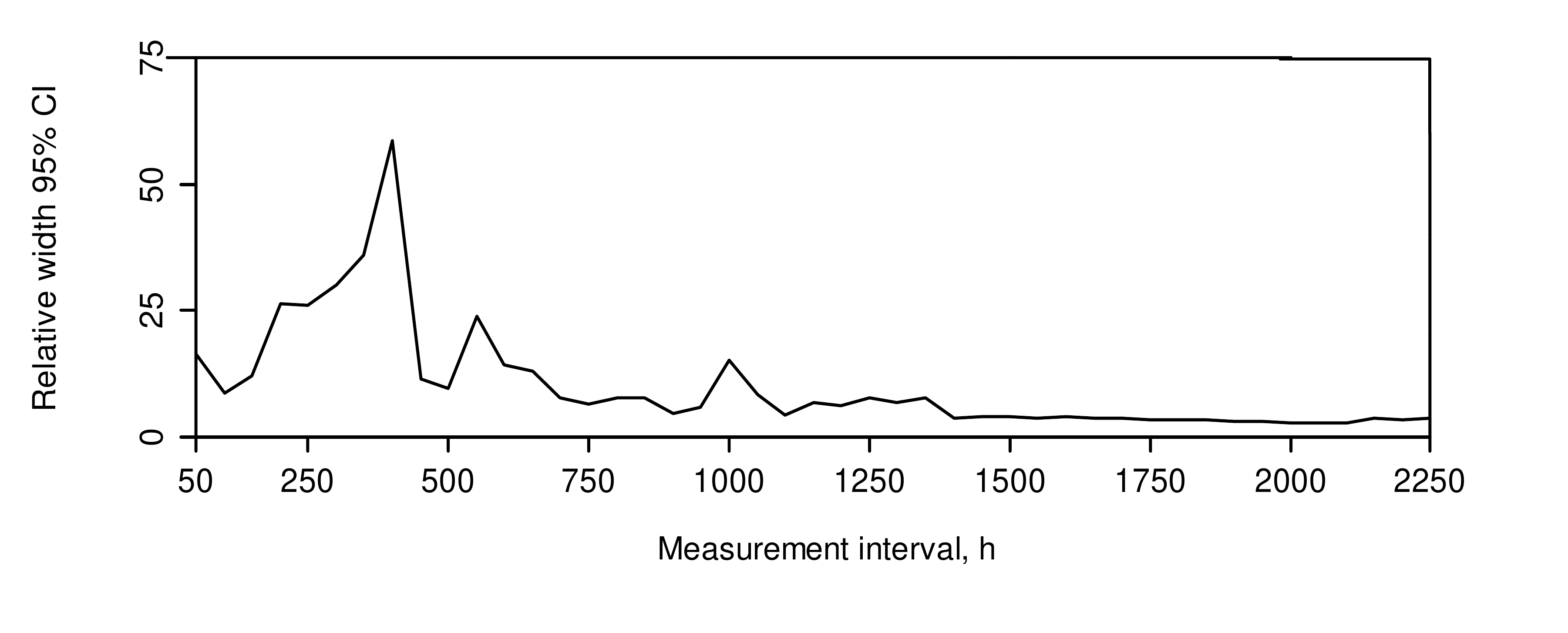}
\caption[Estimate precision as a function of the measurement interval.]{Estimate precision as a function of the measurement interval.  The estimate precision is the sum of the relative widths of the 95\% confidence interval (CI) for the estimated coefficients of the network model.  The CI are calculated empirically using the normal approximation from the 250 (in the simple case) and 100 (in the complex case) replications, respectively.  The coefficients are estimated using the STERGM package \citep{Krivitsky2014STERGM} of the \emph{statnet} suite \citep{statnet} from two network snapshots spaced $h$ time units apart.  The top panel is for the simple dynamic network and the bottom panel for the complex dynamic network.  For the simple dynamic network, the dashed vertical line at $h^*$ shows the analytically derived optimum measurement interval (see section \ref{sec2.2}).}
\label{fig5}
\end{figure}

Figure~\ref{fig5} shows the objective function (i.e., the sum of the relative width of the CIs of the coefficient estimates) for both the simple and complex dynamic network.  In the simple case we see that a measurement interval given by the approximation for $h^*$ gives close to optimal performance.  A precise value for the optimal measurement interval is hard to determine in this figure because the lines are quite jagged due to our small number of replications.  Luckily, similar to what we saw in the theory section (specifically section~\ref{sec2.2}), the objective function has a wide and flat bottom making many choices of the measurement interval close to optimal.  We expect that if we simulated longer and took even longer measurement intervals, the objective function would curve back up.  Both the number of replications and the simulation time were limited by our available processing time.

\section{Discussion}\label{sec4}
We solve the problem of determining the optimal time interval, $h^*$, between two cross-sectional network observations when designing a longitudinal network study, by placing it into a statistical framework. Specifically, we choose $h^*$ to minimize the sum of the relative widths of the 95\% confidence intervals of the network parameters we are estimating.  This is our first contribution.  We demonstrate the practicality of this approach by using it to evaluate a range of timings for a hypothetical longitudinal network study.  Specifically, we simulated a dynamic network and then estimated the parameters, repeatedly in order to determine the expected size of the CI for different measurement spacings.  Both the dynamic network model (longitudinal ERGM \citep{Snijders2013}) and the estimation procedure (STERGM \cite{Krivitsky2014STERGM}) are common tools of social network researchers.

Our second contribution is the simple approximation for $h^*$ for the case where each parameters corresponds to the rate for a different network timers (e.g., edge formation, edge dissolution, infection, etc.).  In particular, writing it as a weighted sum \eqref{eq:hstar2} we saw that $h^*$ is mainly determined by those timers which occur infrequently and have short expected durations.  This makes intuitive sense, since the parameters for infrequent timers have greater uncertainty and for any absolute error the relative error is greater for small values.  In our example, of iid edge formation and dissolution (section~\ref{sec2.3}), the dissolution timers are generally fewer than the formation timers and have shorter durations (assuming the average degree is much less than the number of nodes), implying that $h^*$ is close to a small multiple of the expected tie duration, $\gamma_0/d$.  Our computational experiments for the simple dynamic network suggest that a measurement interval somewhat greater than the approximation for $h^*$ is often slightly better.
Our theoretical analysis assumed that the various rates are independent and thus, extending the theoretical analysis to the dynamic networks we used in our computational experiments (the dynamic ERGM simulations using the algorithm of \citet{Snijders2013}) is an area for future research.

A third, minor contribution is the observation both in the theory and the computational sections that objective function is quite flat near $h^*$.  This means that there is a wide range of values with relative confidence intervals that are almost minimal and that designers of a longitudinal study can incorporate other (e.g., feasibility) considerations when choosing $h$, without sacrificing much precision of the resulting estimates.  One technical consideration that should influence our choice of $h$ is the assumption we made here that that no node pair changes state more than once in that time period, or more generally that no secondary events occur in the measurement period (e.g., that a newly infected person infects other people before the next cross-sectional network observations).  This assumption suggests that we should choose an $h$ smaller than the expected time of the first few secondary events.  These details are areas of future research.

One limitation on the applicability of our framework is that longitudinal studies often have more than snapshot information. Social network surveys, may for example ask when a relationship formed or how long it lasted.  Extending our framework to incorporate such information could be one area of future work as well as an extension of our framework to clinical studies in the fields of network epidemiology where both clinical data and social network data are sampled such as a longitudinal HIV-focused cohort study of young men who have sex with men (YMSM) in Chicago \citep{Kuhns2015,Mustanski2014,Mustanski2015}). Here, both clinical processes such as infection processes and disease progression as well as social network processes have to be considered to allow for an overall optimal timing between measurements.  Another area for future work is to apply this framework to the tactical decisions of which individual to survey next rather than the strategic decision of the spacing survey waves.

\textbf{Acknowledgements.} We thank Peter Baumann, Michelle Birkett, Noshir Contractor, Alexander Gutfraind, and Brian Mustanski for helpful comments.

\clearpage
\bibliography{refOptimal}

\section{Appendix}

\begin{figure}[h]
\centering
\includegraphics[width=\textwidth]{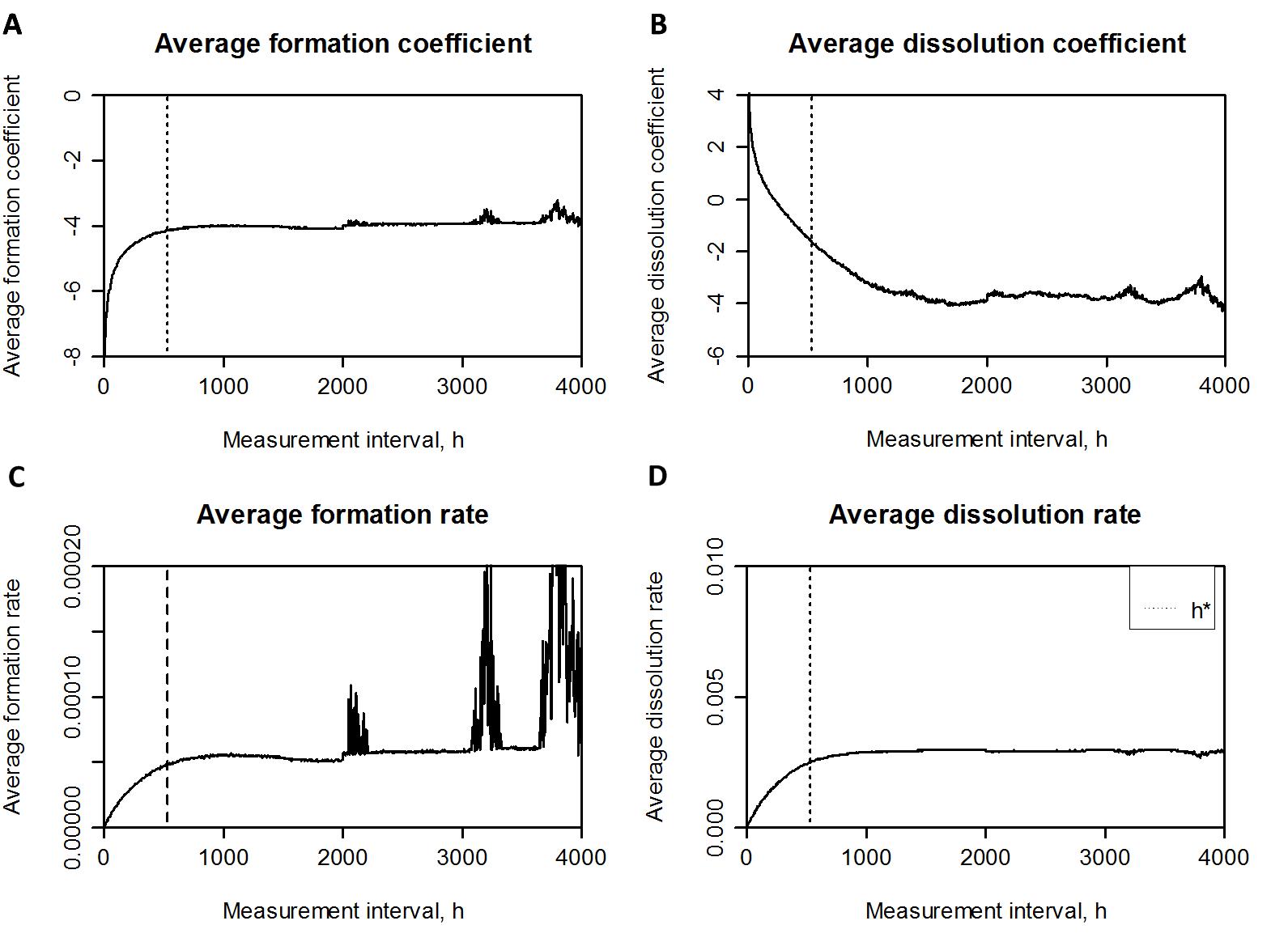}
\caption[Average estimates of formation and dissolution for the simple dynamic network]{Estimates of formation and dissolution rates and coefficients for the simple dynamic network averaged over 250 replications for different measurement intervals.  Coefficients were estimated using the STERGM package \citep{Krivitsky2014STERGM} of the \emph{statnet} suite \citep{statnet} from two network snapshots spaced $h$ time units apart.  The formation and dissolution rates were derived from the coefficients according to \citet{Snijders2013}.  The dashed vertical line at $h^*$ shows the analytically derived optimum measurement interval (see section \ref{sec2.2}).}
\label{fig:appendix}
\end{figure}

\end{document}